\documentclass[12pt,preprint]{aastex}
\slugcomment{LA-UR-06-6066}
\shorttitle{$\rm Ly \, \alpha$ in cool white dwarf atmospheres}
\shortauthors{Kowalski \& Saumon}

\def\Teff{T_{\rm eff}}
\def\teff{T_{\rm eff}}

\def\wig#1{\mathrel{\hbox{\hbox to 0pt{%
          \lower.5ex\hbox{$\sim$}\hss}\raise.4ex\hbox{$#1$}}}}

\begin{document}

\title{Found: The missing blue opacity in atmosphere models of cool hydrogen white dwarfs}
\author{P. M. Kowalski\altaffilmark{1,2} \& D. Saumon\altaffilmark{2,1}}

\affil{\altaffilmark{1}Department of Physics and Astronomy, Vanderbilt University, Nashville, TN
37235-1807}
\affil{\altaffilmark{2}Los Alamos National Laboratory, MS P365, Los Alamos, NM 87545}
\email{kowalski@lanl.gov}
\email{dsaumon@lanl.gov}

\begin{abstract}

We investigate the importance of the far red wing of the Lyman $\,\alpha$ line of hydrogen in the 
atmospheres of cool white dwarfs of pure hydrogen composition. We find that this absorption process 
dominates all important sources of opacity in the blue part of the optical spectrum
of these stars. Our successful fits to the spectra of cool DA/DC white dwarfs
indicate that the far red wing of the $\rm Ly \, \alpha$ line is the source of opacity that 
had been missing in the models.
The observed sequence of cool white dwarfs in color-color diagrams is very well reproduced by our 
new pure hydrogen atmosphere models, suggesting that the atmospheric composition of
the coolest DC white dwarfs must be revisited.

\end{abstract}


\section{Introduction}

The detection and analysis of large samples of cool white dwarfs
over the past decade has led to a significant improvement in our
knowledge of the chemical composition and evolution of their
atmospheres \citep{Har06, Kilic06, BLR01,
BRL97}. The difficulties in interpreting the spectra of some
very cool white dwarfs thought to posses helium-rich atmospheres
with small amounts of H \citep{Bergeron05,BL02,OP01} are most probably related to the
extreme physical conditions found in their atmospheres
\citep{Kowalski06a}. On the other hand, the physical conditions in
hydrogen-rich atmospheres are much less extreme and the current
models should be much more reliable. Nonetheless, models systematically
overestimate the flux for $\lambda \wig<5000 \,$\AA\ of
hydrogen-rich white dwarfs of $T_{\rm eff}\rm <6000 \, K$
\citep{BLR01,Bergeron01,BRL97}. \citet{BRL97}
suggested that the mechanism responsible for this phenomenon could
be the pseudo-continuum bound-free opacity from hydrogen atoms
perturbed by collisions in a relatively dense medium. However, a realistic calculation of this
opacity source shows that it is too weak to correct the
observed blue flux excess in the models \citep{Kowalski06b}.

In this paper we consider the possibility that the missing opacity is the red wing of the 
pressure broadened hydrogen Lyman$\, \alpha$ line.  This absorption mechanism results
from the perturbations of hydrogen atoms by their interaction with other particles, primarily $\rm H$ and $\rm H_2$.
Such perturbations result in the lowering of the $\rm Ly \, \alpha$ transition energy
and the possibility of a bound-bound transition for photon energies that are smaller than that of
an isolated hydrogen atom ($E_{\rm 12}^{\rm 0}\rm =10.2 \, eV$).
We present a new, semi-classical calculation of this process based on accurate quantum mechanical 
calculations and a model that has no free parameters.
The inclusion of broadening by collisions with $\rm H_2$ is the key element of this 
calculation. The resulting pure hydrogen models successfully reproduce
the spectral energy distributions (SED) and colors of very cool DA and DC white dwarfs.

\begin{figure}[t]
\epsscale{0.7}
\plotone{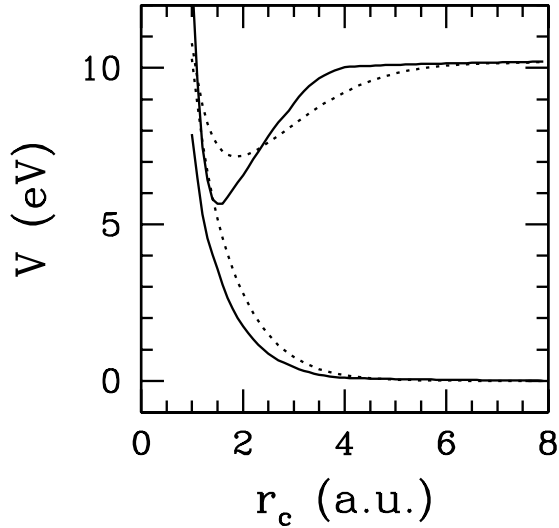}
\figcaption[f1.ps]{
The interaction energy curves as a function of the inter-particle collision distance for the 
ground state (lower curves) and the first excited Rydberg
state (upper curves) of $\rm H-H_2$ (solid) and $\rm H-H$ (dotted) dimers.
For $\rm H-H_2$ the energy curves shown are for a collision angle
of $\rm 90^{o}$, where the collision angle is defined between the line connecting 
the $\rm H$ atom to the center of $\rm H_2$ and the molecular axis.
       \label{F1}}
\end{figure}

\section{Theoretical approach}

We are interested in the $\rm Ly \, \alpha$ opacity far from the
line center, at wavelengths $\lambda\wig>2500 \,$\AA. A
lowering of the $\rm Ly \, \alpha$ transition energy by more than
$\sim 5 \, \rm eV$ is required for a bound-bound absorption from the
ground state of hydrogen atom at these wavelengths. This can occur
in rare, close-range collisions that strongly perturb the bound states of the atom.
Because the ground state interactions are strongly repulsive at short range, the
probability of such close-range collisions is much smaller than unity.
Multi-particle collisions that result
in a similar decrease in the $\rm Ly \, \alpha$ transition energy are therefore
insignificant \citep{Kowalski06b} and we
consider the interaction between a $\rm H$ atom and its closest
neighbor only. This approach was used to successfully explain the
complex shape of the Lyman $\rm \alpha$ line wings detected in the
spectra of white dwarfs with $T_{\rm eff}\rm \sim 12000 \, K$
\citep{Allard04}, the UV flux deficiency observed in the white dwarf star
$\rm L 745-46A$ \citep{Koester00}, and to model resonance broadening of the $\rm Ly \, \alpha$ line
in the solar spectrum \citep{SD69}.

For a given colliding pair, the change in the $\rm Ly \, \alpha$ transition energy results 
from the formation of a temporary dimer, whose first transition energy
$E_{\rm 12}$ differs from that of an isolated hydrogen atom, $E_{\rm 12}^{\rm 0}$. 
$E_{\rm 12}$ is given by the differences in the energies of
the ground state and first excited Rydberg state of a $\rm H$-perturber dimer calculated at a 
fixed inter-particle separation $r_{c}$: $E_{12}(r_c)=E_2(r_c)-E_1(r_c)=h\nu_{12}$,
where $\nu_{12}$ is the frequency of the absorbed photon.
This picture is in the spirit of the Franck-Condon principle, which states that
the nuclei remain in fixed positions during a radiative transition \citep{Davydov}. 
The differential probability of finding such a dimer with an inter-particle
separation between $r_c$ and $r_c+dr_c$ is given by \citep{MS}
\begin{equation} dP_{c}(r_c)=n_{pert}r_c^2dr_c\left
   (\int_{\theta,\phi}{e^{-V_{\rm H-pert}(r_c,\theta,\phi)/k_BT} \sin \theta d \theta d \phi}\right), \label{11}
\end{equation}
where $n_{pert}$ is a number density of perturbers, $V_{\rm H-pert}$ is the interaction energy 
between a hydrogen atom and the perturber localized at the position $(r_c,\theta,\phi)$
in relation to the hydrogen atom. The potential for the interaction of a hydrogen atom
with H ($\rm H - H$ dimer) is from \citet{KW} and that for perturbations by H$_2$ ($\rm H-H_2$ dimer)
is from \citet{Bo96,BOO} (Fig. 1). 
For the $\rm H-H$ dimer, the allowed bound-bound transition from the ground state
with the smallest transition energy (i.e. greatest broadening) at all values of $r_c$ is 
from ${\rm b}\, ^{3}\Sigma^{+}_{u} \, 2p\sigma$ to ${\rm a} \,^{3}\Sigma^{+}_{g} \, 2s\sigma$.
Satellite features \citep{Allard94} do not appear for this transition because the difference between
the two energy curves has no extrema (Fig. 1). 
Contributions from transitions to other excited levels of 
the $\rm H-H$ dimer (such as ${\rm h} \,^{3}\Sigma^{+}_{g} \, 3s\sigma$  and ${\rm i} \,^{3}\Pi_{g} \, 3d\pi$)
are negligible in the far red wing ($\lambda \wig>2000\,$\AA) because they
always lie above the ${\rm a}\,^{3}\Sigma^{+}_{g}$ state by at least 2.5$\,$eV in the relevant range of $r_c$
and hence only provide absorption for photons with higher energies than the ${\rm a}\,^{3}\Sigma^{+}_{g}$ state.  
For the same reason, transitions from the binding $\rm H-H$ ground state 
($X\, ^{1}\Sigma^{+}_{g}(1s\sigma)^{2}$) are also negligible in the far red wing.
We use the $\rm a\, ^{3}\Sigma^{+}_{g} \, 2s\sigma$ energy curve from \citet{SW99}.
For the energy of the first excited Rydberg state of the $\rm H-H_2$ system, we used 
the H$_3$ calculations of \citet{Bo96}.  To obtain the H -- H$_2$ potential from the H$_3$
potential surface, we assumed a fixed internuclear separation for $\rm H_2$ 
of $\rm 1.4 \, a.u.$ This approach is justified by the fact
that the atoms in $\rm H_2$ vibrating in the ground state  spend most of the 
time at the equilibrium internuclear separation.  At the temperatures of interest 
($\rm T<6000\,$K), the vibrational excitation is small, and so is the amplitude of vibration ($\rm \wig< 0.4\,a.u.$).
We have verified that such a small change in the intermolecular separation does not affect our results significantly.

The line profile is calculated in the quasi-static approximation, which is suitable for a calculation of 
the far line wings perturbed by classical particles \citep{AK82}. The profile of the line $\alpha(\nu)$ 
is given by (\citet{AK82}, Eqs. 13 \& 61):
\begin{equation} 
\alpha(\nu)d\nu \sim dP_c(r_c(\nu)) |\langle2|\vec{d}|1\rangle|^2 \rm , 
\end{equation}
where $\langle2|\vec{d}|1\rangle$ is the $r_c$-dependent transition dipole moment 
between the ground state and the first excited Rydberg state of the colliding pair, and $\nu$ is the frequency.
The dipole transition moments, which vary with separation $r_c$, are from \citet{SW99} for $\rm H-H$ and 
\citet{Pe95} for $\rm H-H_2$.
For a given $\rm Ly \, \alpha$ transition energy,
the $\rm H-H_2$ ground state interaction is much less repulsive than the $\rm H-H$ ground state interaction 
(Fig. 1), and the probability of a close $\rm H-H_2$ interaction is much larger than that of a close $\rm H-H$ 
interaction ($dP_c(\rm H-H_2 \it )>>dP_c(\rm H-H \it )$). Hence, the $\rm H-H_2$ collisions
are the dominant contributors to the $\rm Ly \, \alpha$ line red wing opacity in the atmospheres of 
cool H-rich white dwarfs with a significant abundance of $\rm H_2$ ($T_{\rm eff}\rm<6000 \, K$).

\section{Applications to cool white dwarf atmospheres}

\subsection{The spectral energy distribution  of cool DAs}

\begin{figure}[ht]
\epsscale{0.6}
\plotone{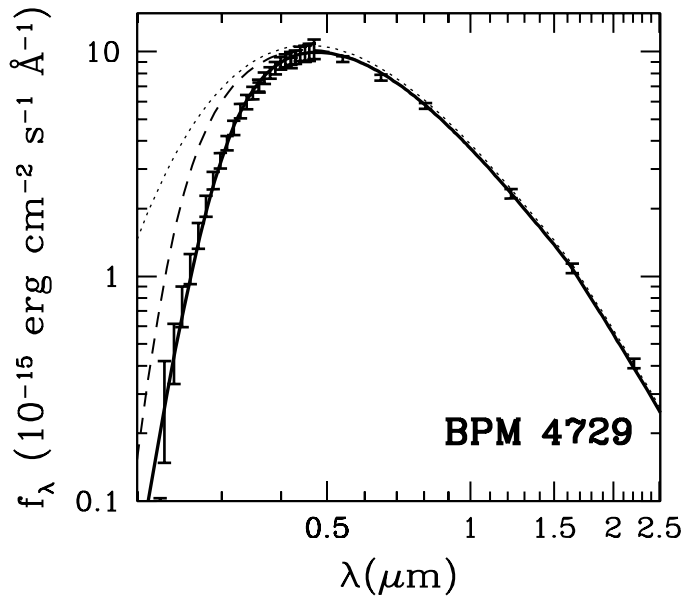}
\figcaption[f2.ps]{
The spectral energy distribution of the DA white dwarf BPM 4729 (WD 0752$-$676). The UV spectrum 
of \citet{Wolff02} extends up to 4500$\,$\AA.  Additional measurements are
broadband fluxes from U \citep{MS99} and BVRIJHK \citep{Bergeron01} photometry. 
The solid and dotted lines represent the pure hydrogen models with and without 
the opacity from the red wing of the Ly$\, \alpha$ line, respectively.  The fit parameters are
$\Teff =5820 \,$K and $\log g = 8.30$. 
The dashed line represent the spectrum obtained when only $\rm H-H$ collisions are considered 
in the Ly$\,\alpha$ opacity calculation.  All model spectra shown are computed from the same
atmospheric structure.
       \label{F2}}
\end{figure}

\begin{figure}
\epsscale{0.8}
\plotone{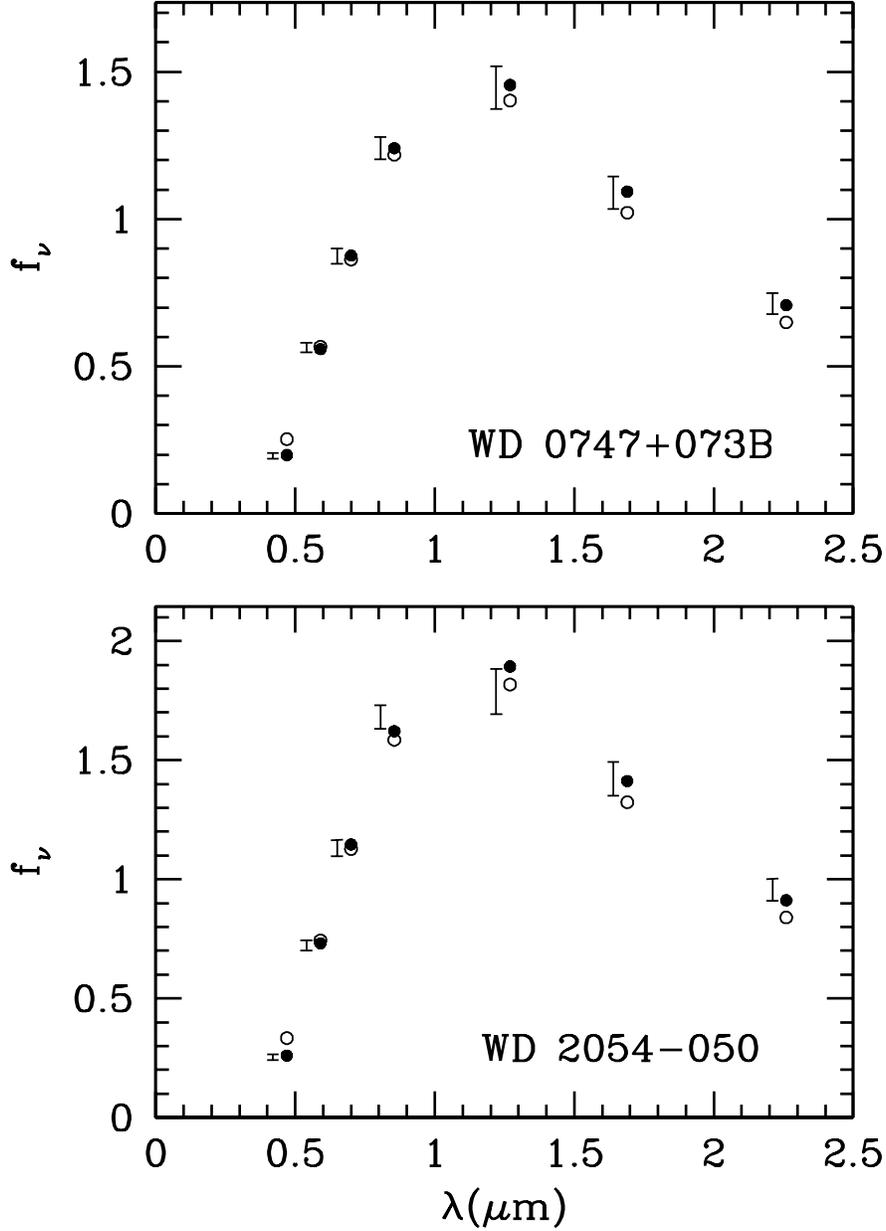}
\figcaption[f3.ps]{
Fits to the spectral energy distributions of two very cool white dwarfs from the sample of \cite{Bergeron01}. 
Bars represent the observed BVRIJHK fluxes with $\pm 1\sigma$ errors 
\citep{Bergeron01}. Circles (shifted by $+0.05\, \mu$m for clarity, represent the best fitting pure hydrogen models.  
Models with the Ly$\,\alpha$ opacity
(filled circles) give the following $(\Teff,\,\log g)$: (4255, 7.78) for WD 0747+073B and (4260, 7.84) for
WD 2054$-$050.  Models computed {\it without} the Ly$\,\alpha$ opacity (open circles) give (4240, 7.75) for WD 0747+073B 
and (4260, 7.83) for WD 2054$-$050.  The units of flux are $\rm 10^{-26} \, erg \, cm^{-2} \, s^{-1} \, Hz^{-1}$. 
\citet{BLR01} classified WD 2054$-$050 as a He-rich star.
       \label{F3}}
\end{figure}

BPM 4729 (WD 0752$-$676) is the only cool DA white dwarf observed in the near UV
\citep{Wolff02}. When complemented with the BVRIJHK
photometry of \citet{BLR01} and the U magnitude  \citep{MS99}
the complete spectral energy distribution (SED) of this cool white dwarf is obtained. The UV
spectrum extends well into the wing of the $\rm Ly \, \alpha$ line and thus
provides an excellent measure of the strength and profile of the
line. Our fit of a pure H model  ($T_{\rm eff}=5820 \,$K, $\log
g=8.30$) to the entire SED of BPM 4729 is excellent (Fig. 2). These
values of $T_{\rm eff}$ and $g$ agree with the values of
\citet{BLR01} at the $1\sigma$ level. The high quality of the fit of the UV spectrum
demonstrates the validity of our model for the far red wing of the
$\rm Ly \, \alpha$ line under the conditions encountered in BPM 4729.
At the photosphere, the composition is $\log n({\rm H})=19.8$ and $\log
n({\rm H_2})=19.0$, where $n$ is the number density in $\rm cm^{-3}$. Despite
the lower abundance of $\rm H_2$, it is an important contributor to
the $\rm Ly \, \alpha$ line opacity in the far red wing (Fig. 2).
\citet{Wolff02} obtained a good fit of the UV and blue spectrum with a 
He-rich model with $\rm H/He=3 \times 10^{-5}$.
They point out that such a low H abundance fails to reproduce the Balmer lines of BPM 4729,
however. \citet{Wolff02} required a large amount of $\rm He$ to match the UV spectrum
because the broadening of Ly $\alpha$ by 
collisions with He is weaker than with H or H$_2$.  They also 
considered a pure hydrogen model but its Ly $\alpha$ line extends only up to $\sim 2400\,$\AA.
In view of the difference between the potential curves for the 
${\rm b}\, ^{3}\Sigma^{+}_{u}$ to ${\rm a} \,^{3}\Sigma^{+}_{g}$ states of the $\rm H-H$ dimer 
(Fig. 1) this indicates that this particular transition was not included in the model
of \citet{Wolff02}.
The main reason for our success with a pure H model is our
inclusion of the broadening of $\rm Ly \, \alpha$ line by collisions of hydrogen atoms with $\rm H_2$
and using the  b $\rightarrow$ a transition for the H$-$H dimer.

We have also fitted BVRIJHK photometry of most of the coolest DA/DC white dwarfs in the \citet{BLR01} sample. 
Two typical fits are shown in Fig. 3. Our
models have no difficulty in reproducing the observed B flux and excellent fits are obtained for most   
stars with either pure H or pure He. The importance of the $\rm Ly \, \alpha$ line opacity
in the models is revealed by comparing with fits of models that exclude this opacity source.
The improvement in fitting the B flux with our Ly$\,\alpha$ opacity model is clearly visible.
The star WD $2054-050$ (vB 11, lower panel of Fig. 3) was assigned a pure helium
composition by \citet{BLR01}.  We will return to this point in the next section.
The success of the pure H models indicates that the $\rm Ly \, \alpha$ red wing opacity is
the missing blue opacity source in models of cool H-rich white dwarf atmospheres.

\subsection{Color-color diagrams}

Color-color diagrams allow a broader comparison with data.
Figure 4 shows two large samples of cool WDs, that of
\citet{BLR01}, and a combination of two samples culled from
the Sloan Digital Sky Survey \citep{Kilic06,Har06}. 
The atmospheric composition of the stars in the first sample, 
as determined by \citet{BLR01}, is indicated in the left panel of Fig. 4.
Colors computed from our pure
H model sequence, both with and without the $\rm Ly \, \alpha$ opacity,
are shown.  Our sequence of new pure hydrogen atmospheres
follows the observed sequence of stars very well in these two color-color
diagrams.  Equally good agreement is found in all other color-color diagrams 
involving BVRIJHK and {\it ugriz} colors (not shown here).  

\begin{figure}
\epsscale{1.0}
\plotone{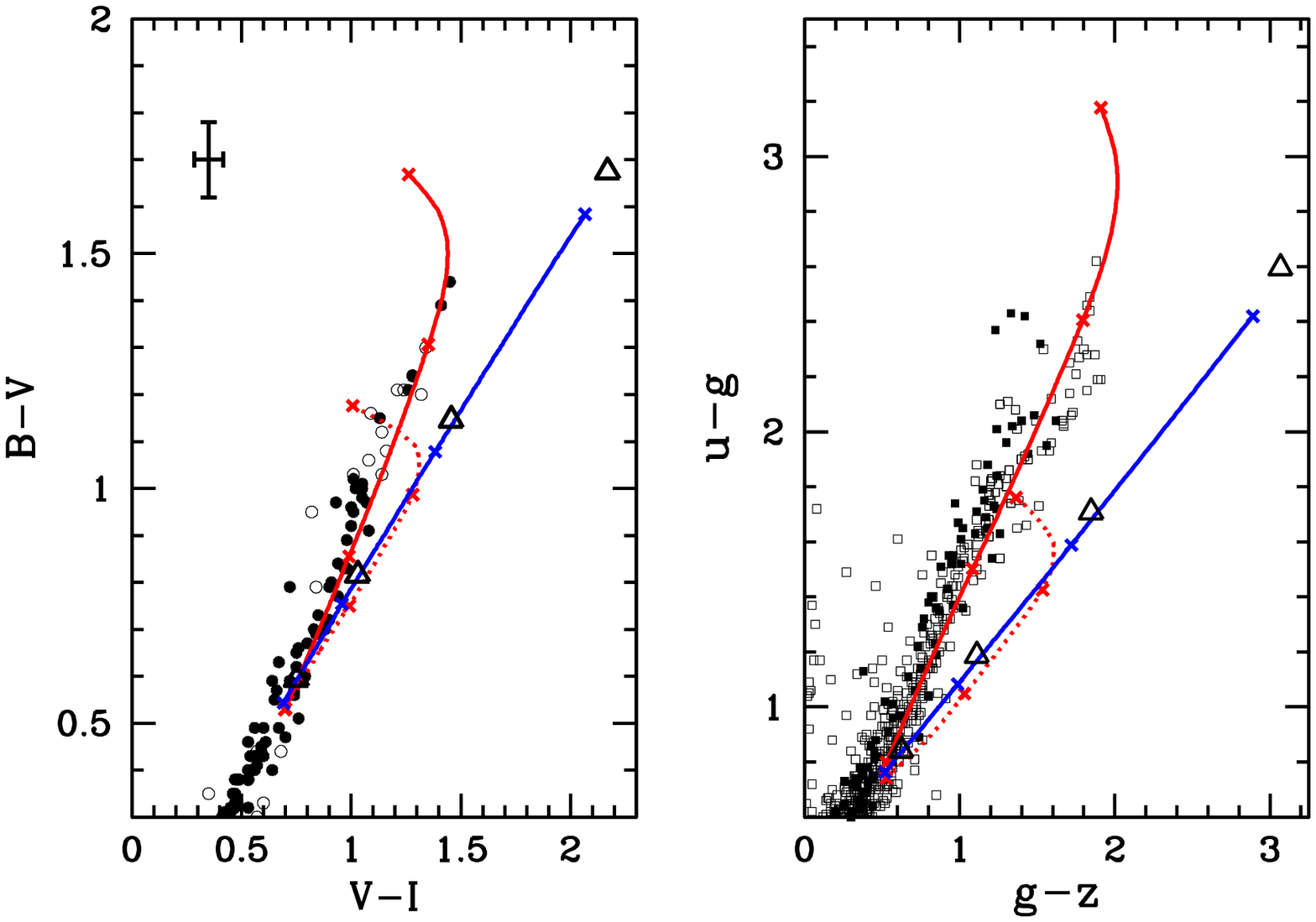}
\figcaption[f4.ps]{Color-color diagrams for cool white dwarfs. {\it Left panel:} The sample of \citet{Bergeron01}. 
The composition as determined by \citet{Bergeron01} is shown by filled circles (hydrogen-rich) and open 
circles (helium-rich) and the photometric uncertainties are shown by the error bars in the upper left. 
{\it Right panel:} the SDSS white dwarfs sample of \citet{Kilic06} (filled squares) 
and \citet{Har06} (open squares).  The photometric error bars vary widely from star to star
in these two samples and are not shown.  The lines represent the synthetic colors of the models: 
pure hydrogen with the Ly$\,\alpha$ line opacity (red, solid) and without (red, dotted).  A pure He sequence is also
shown (blue, dashed lines).  The $T_{\rm eff}$ along each curve is indicated by crosses at 3000, 4000, 5000, 6000$\,$K 
from top to bottom. All models have $\log g = 8$. The corresponding colors of black bodies  are shown by
open triangles.  DQ, DZ and known or suspected double degenerates stars 
have been removed from the samples.  
       \label{F4}}
\end{figure}

On the other hand, the sequence of pure 
helium models of \citet{bsw95} (see also \citet{bwb95})
used in the analysis of \citet{BRL97} and \citet{BLR01} also reproduces the observed sequence in the BVI 
color-color diagram.  This
potential ambiguity in the atmospheric composition of cool DC stars is resolved
with a new sequence of pure He models.  We have computed pure He models based on much 
improved constitutive physics for dense helium, including a better calculation
of the ionization equilibrium and of the He$^-$ ff and Rayleigh scattering
absorptions \citep{irs02,KS04,KSM05,KMS06}.  A detailed description of these
models will be the subject of a future publication.  Qualitatively, the correlations
in dense fluid He reduces the contribution of Raleigh scattering by a factor of
$\sim 10$ \citep{irs02} and the strong interactions in the fluid increase the
ionization fraction (and hence, the He$^-$ ff opacity) by 2-3 orders of magnitude
\citep{KSM05,KMS06}.  The combination of these effects make He$^-$ ff the only 
important source of opacity in these models. On the other hand, because of a lower
ionization fraction and the higher Rayleigh scattering opacity in a dilute gas,
the pure He model spectra of \citet{bsw95}  are affected by both opacities (see
their Fig. 13).
Since He$^-$ ff is a nearly gray opacity, our pure He atmosphere models are essentially gray
and the emergent flux is close to that of a black body.  The colors of this pure
He sequence are shown in Fig. 4 along with those of black bodies.
Compared to the \citet{bsw95} pure He sequence,
to our pure H sequence, and to the observed
sequence of very cool WDs, this new pure He sequence is much redder for $\teff\wig< 4500\,$K. 
As the ionization fraction is sufficiently large for the He$^-$ ff opacity to dominate, the colors of
our new He sequence are insensitive to modest pollution by metals since increasing
the fraction of free electrons will only increase the He$^-$ ff opacity without
any effect on the emergent spectrum.
On the basis of the location of the pure H and pure He model
sequences in the color-color diagrams (Fig. 4) and the excellent fit
we obtain for WD2054$-$050 with a pure hydrogen model, it appears that the
atmospheric composition of the coolest DC stars needs to be revisited.

\section{Conclusions}

The existence of an unidentified absorption mechanism in cool hydrogen
white dwarfs atmospheres was reported a decade ago
\citep{BRL97}. The interpretation of this missing opacity as
the pseudo-continuum absorption from hydrogen atoms has been
recently shown to be incorrect \citep{Kowalski06b}. On the other
hand, the red wing of the $\rm Ly \, \alpha$ line opacity from hydrogen
could provide the required absorption \citep{Wolff02}. We present a
new calculation of the extreme pressure-broadening of the $\rm Ly \,
\alpha$ line by both $\rm H$ and $\rm H_2$. When included in our new
pure hydrogen atmosphere models, we obtained an excellent agreement
with the UV/optical/near-IR spectral energy distribution (SED) of the cool 
DA white dwarf BPM 4729
and we successfully fitted the B through K spectral energy
distributions of stars with hydrogen-rich atmospheres, as
determined by \citet{BRL97} and \citet{BLR01}.
The inclusion of broadening by collisions with H$_2$ is essential to 
reproduce the SED of very cool hydrogen white dwarfs.

In color-color diagrams,
the new pure hydrogen models follow the observed sequences of cool
white dwarfs very well, while an improved physical description of dense
helium moves the cool end of the helium sequence to the red, near the locus of black body colors.
This suggest that detailed fitting of the SED of the coolest DC white dwarfs
will result in a number of them, and perhaps most, having hydrogen-rich
composition rather than helium-rich.  An example is provided by our fit to
WD$2054-050$.  This has direct implications for the spectral evolution of
very cool white dwarfs and for the physical mechanisms responsible for their
atmospheric composition.

This research was supported by the United States Department of Energy under contract W-7405-ENG-36.
We thank P. Bergeron for illuminating discussions.


\end{document}